\documentclass{article}
\usepackage{amsmath}
\usepackage[ruled,linesnumbered]{algorithm2e}
\SetKwComment{Comment}{/* }{ */}
\usepackage{amsthm}
\usepackage{placeins} 
\usepackage{tikz} 
\usetikzlibrary{positioning, 
                quotes}

\newtheorem{theorem}{Theorem}

\newcommand{\calS}{\mathcal{S}}
\newcommand{\cchoice}{c_{\mathrm{choice}}}
\newcommand{\cunify}{c_{\mathrm{unify}}}
\newcommand{\s}[1]{\textsf{#1}}

\newcommand{\fr}{\beta}
\newcommand{\pr}{\mathit{part}}
\newcommand{\com}{\mathit{com}}
\newcommand{\unc}{\mathit{unc}}
\newcommand{\last}{\mathit{last}}
\newcommand{\lastj}{\mathit{lastj}}

\title{A faster algorithm for the construction of optimal factoring automata}
\author{Thomas Erlebach \and Kleitos Papadopoulos}
\begin{document}
\maketitle
\begin{abstract}
The problem of constructing optimal factoring automata arises in the context
of unification factoring for the efficient execution of logic programs.
Given an ordered set of $n$ strings of length~$m$, the problem is to construct
a trie-like tree structure of minimum size in which the leaves in left-to-right
order represent the input strings in the given order.
Contrary to standard tries, the order
in which the characters of a string are encountered can be different on different
root-to-leaf paths. Dawson et al.\ [ACM Trans.\ Program.\ Lang.\ Syst.\ 18(5):528--563, 1996]
gave an algorithm that solves the problem in time
$O(n^2 m (n+m))$.
In this paper, we present an improved algorithm with running-time
$O(n^2m)$.
\end{abstract}

\section{Introduction}
The execution of programs written in a logic programming language
such as Prolog relies on unification as the basic computational
mechanism. A Prolog program consists of a set of rules, and the
system needs to match the head of the goal with the head of each
of the rules that can be unified with the goal. Therefore,
preprocessing the rule heads in order to speed up the unification process,
called \emph{unification factoring},
is important. Dawson et al.~\cite{dawson1996principles} describe
how this preprocessing problem translates into the problem of constructing
an optimal factoring automaton. This problem can be viewed as the purely combinatorial
problem of computing a certain trie-like tree structure of minimum size for a given
ordered set of strings.
We focus on this combinatorial problem in
the remainder of the paper and
refer to~\cite{dawson1996principles} for
the relevant background on logic programming and for the details of how the preprocessing
problem for the rule heads translates into the problem of constructing optimal
factoring automata.
Unification factoring has been successfully implemented in at least one Prolog
system, namely XSB~\cite{rao1997xsb}.

\begin{figure}
\centering
\begin{tikzpicture}[nodes={circle}, -]
  \node[draw] (r) at (0,6) {$r$};
  \draw (0,6) node[right=0.15] {$1$};
  \node[draw] (u) at (-4,4) {$u$};
  \draw (-4,4) node[right=0.15] {$2$};
  \node[draw] (v) at (0,4) {$v$};
  \draw (0,4) node[right=0.15] {$2$};
  \node[draw] (w) at (4,4) {$w$};
  \draw (4,4) node[right=0.15] {$3$};
  \node[draw] (x) at (-4,2) {$x$};
  \draw (-4,2) node[right=0.15] {$3$};
  \node[draw] (y) at (0,2) {$y$};
  \draw (0,2) node[right=0.15] {$3$};
  \node[draw] (z) at (4,2) {$z$};
  \draw (4,2) node[right=0.15] {$2$};
  \node[draw,rectangle] (aaa) at (-4,0) {\s{aaa}};
  \node[draw,rectangle] (bbc) at (0,0) {\s{bbc}};
  \node[draw,rectangle] (aab) at (3,0) {\s{aab}};
  \node[draw,rectangle] (acb) at (5,0) {\s{acb}};

  \draw (r) edge["\s{a}"] (u);
  \draw (r) edge["\s{b}"] (v);
  \draw (r) edge["\s{a}"] (w);
  \draw (u) edge["\s{a}"] (x);
  \draw (v) edge["\s{b}"] (y);
  \draw (w) edge["\s{b}"] (z);
  \draw (x) edge["\s{a}"] (aaa);
  \draw (y) edge["\s{c}"] (bbc);
  \draw (z) edge["\s{a}"] (aab);
  \draw (z) edge["\s{c}"] (acb);
\end{tikzpicture}

\caption{Optimal FA for the string tuple $\calS=(\s{aaa},\s{bbc},\s{aab},\s{acb})$. For each non-leaf node $q$, the position $p(q)$ is shown to the right of the node.}
 \label{fig:FA}
\end{figure}
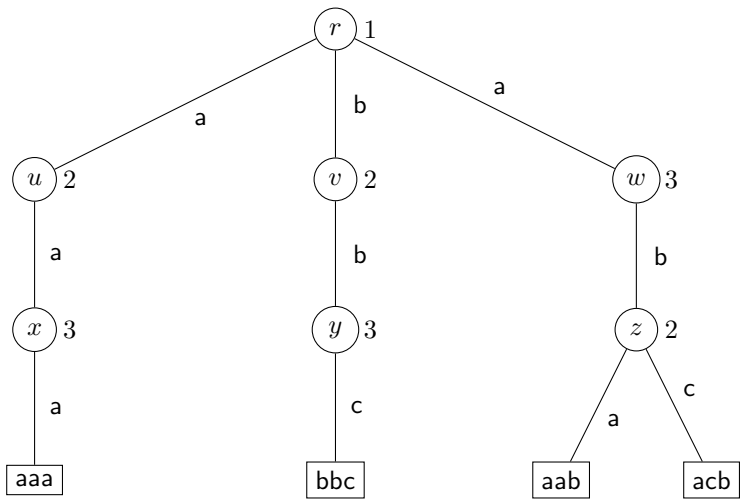
The trie-like tree structures of interest can be defined as follows.
Given an ordered set $\calS$ of $n$ strings $(S_1,S_2,\ldots,S_n)$ of
equal length $m\ge 1$ over some alphabet~$\Sigma$,
a \emph{factoring automaton} (FA) is a
rooted tree $T=(V,E)$ with $n$ leaves, all of which have depth~$m$.
Each non-leaf node $v$ is labeled with a position $p(v) \in [m]$,
and each edge is labeled with a character from~$\Sigma$.
For an edge $e$ between a non-leaf node $v$ and a child of~$v$,
we say that $v$ is the \emph{parent node} of~$e$.
On each root-to-leaf path, every position in $[m]$ occurs as the label
of a non-leaf node exactly once. For every non-leaf node, the edges to its
children are ordered, and the labels of any two consecutive such edges are
different (but it is possible that the labels of two edges that are not
consecutive are the same).
The path from the root to the $i$-th leaf, for $i\in [n]$, must produce
the string $S_i$. Here, a path $P$ produces the string of length $m$ whose
character in position~$j$, for $j\in [m]$, is equal to the label of the
edge between the node $v$ on $P$ that has label $p(v)=j$ and the child
node of $v$ on~$P$.
We say that the $i$-th leaf \emph{represents} the string~$S_i$, and we may refer to the $i$-th leaf simply as the leaf~$S_i$.
Figure~\ref{fig:FA} shows an example of a FA
for the ordered set of strings, or string tuple,
$\calS=(\s{aaa},\s{bbc},\s{aab},\s{acb})$.

The \emph{size} of a FA is the number of edges. For a given ordered
set of strings of equal length, the goal of the
optimal factoring automaton (OFA) problem is to compute a FA of minimum
size. The FA shown in Fig.~\ref{fig:FA} has size $10$ and is in fact
optimal for the given string tuple of that example.
Dawson et al.~\cite{dawson1996principles} show that
the OFA problem can be solved in $O(n^2m(n+m))$ time
using dynamic programming. The problem and solution approach are also
featured in Skiena's algorithms textbook as a `War Story' illustrating how dynamic
programming can solve practical problems~\cite[Section~10.1]{skiena2020book}.
In this paper, we present an improved algorithm that solves the problem in
$O(n^2m)$ time, which is better than the previously known time bound
by a factor of~$n+m$. Furthermore, we show that our algorithm
can be implemented using $O(n^2+nm)$ space, which is the
same as the space used by the algorithm by
Dawson et al.~\cite{dawson1996principles}.

Dawson et al.~\cite{dawson1996principles} also consider a weighted
variant of the OFA problem in which the cost of a non-leaf node with label $k$
that has at least two children
is $\cchoice(k)$ and the cost of an edge with label $c$ whose parent node $v$ has
label $p(v)=k$ is $\cunify(k,c)$, where $\cchoice$ and $\cunify$ are two non-negative cost functions
given as part of the input. The goal of the weighted OFA problem is to compute
a FA of minimum total cost. Dawson et al.~\cite{dawson1996principles} show that
their algorithm for the unweighted OFA problem extends to the weighted case.
We show that the same holds for our faster algorithm.

We remark that the variant of the OFA problem where there is no restriction
on the order in which the given strings are represented by the leaves (and where
the labels of the downward edges incident with a non-leaf node
must be pairwise different) has been shown to be NP-complete
by Comer and Sethi~\cite{comer1977complexity}.

The remainder of the paper is organized as follows. Section~\ref{sec:prelim}
introduces relevant notation and definitions. Then, in
Section~\ref{sec:dawson}, we briefly recall the original
algorithm by Dawson et al.~\cite{dawson1996principles}.
In Section~\ref{sec:algo},
we present our faster algorithm for the OFA problem and its
adaptation to the weighted OFA problem.
We present our conclusions in Section~\ref{sec:conc}.

\section{Preliminaries}
\label{sec:prelim}
For any natural number $x$ we write $[x]$ for the set $\{1,2,3,\ldots,x\}$.
Let $\calS=(S_1,S_2,\ldots,S_n)$ be a string tuple with $n$ strings of length~$m$.
For $1\le i\le n$ and $1\le j\le m$, we use $S_i[j]$ to denote the character in position $j$ of the string~$S_i$.
For $1\le i\le i'\le n$, we write
$\calS[i,i']$ for $(S_i,S_{i+1},\ldots,S[i'])$, the tuple of consecutive
strings in $\calS$ starting with the $i$-th and ending with the $i'$-th.
We call any such $\calS[i,i']$ a \emph{subtuple} of~$\mathcal{S}$.

For $1\le i\le i'\le n$, we denote by $\com(i,i')$ the set of positions
with the property that all strings in $\calS[i,i']$ have the same character
in that position.
For our example from Fig.~\ref{fig:FA} with $\calS=(\s{aaa},\s{bbc},\s{aab},\s{acb})$,
we have $\com(1,4)=\emptyset$ and $\com(3,4)=\{1,3\}$.
We denote the complement of $\com(i,i')$ by
$\unc(i,i') = [m] \setminus \com(i,i')$.
The function name $\com$ is motivated by
the fact that $\com(i,i')$ represent the positions where all strings
in $\calS[i,i']$ have the same letter in \emph{common},
and the name $\unc$ is short for \emph{uncommon} and represents
the complement.
For $1\le i\le j\le j'\le i'\le n$, define
$\Delta(i,i',j,j')=\com(j,j')\setminus \com(i,i')$.
Intuitively, $\Delta(i,i',j,j')$ denotes the positions where all
strings in $\calS[j,j']$ have the same character but not all strings
in $\calS[i,i']$ have the same character.
Observe that $|\Delta(i,i',j,j')|=|\com(j,j')|-|\com(i,i')|$
as $\com(i,i')\subseteq \com(j,j')$.

Let $T=(V,E)$ be a FA for $\calS$.
Each node $q\in V$ is (implicitly) associated with the subtuple $\tau(q)=\calS[\ell_q,r_q]$
of the strings that are represented by leaf descendants of~$q$. For the
root node $r$ of $T$ we have $\tau(r)=\calS$. For the FA shown in Fig.~\ref{fig:FA},
we have $\tau(r)=\calS=\calS[1,4]$ and $\tau(w)=\tau(z)=\calS[3,4]$, for example.

For a node $q\in V$, let $\alpha(q)$ be the set of the positions
$p(s)$ that appear as labels of the strict ancestors $s$ of~$q$
(and hence the empty set if
$q$ is the root of $T$). For the FA shown in Fig.~\ref{fig:FA},
we have $\alpha(r)=\emptyset$, $\alpha(w)=\{1\}$, and $\alpha(z)=\{1,3\}$, for example.
Let $\fr(q)=[m]\setminus\alpha(q)$ and
note that $p(q)\in \fr(q)$ for every non-leaf node~$v$, as each position
in $[m]$ occurs only once on each root-to-leaf path in~$T$.

For a non-leaf node~$q\in V$ with $\tau(q)=\calS[\ell_q,r_q]$
and any position $j\in \fr(q)$,
the subtuple $\tau(q)$ can be partitioned into subtuples
in such a way that the strings in each subtuple have the same character in position~$j$,
while the strings of consecutive subtuples differ in position~$j$.
We also refer to these subtuples as \emph{runs}.
Let $\pr(\ell_q,r_q,j)$ denote the ordered set of these runs, each represented as
a triple $(i,i',c)$ with $\ell_q\le i\le i'\le r_q$ and $c\in \Sigma$, meaning
that the run is $\calS[i,i']$ and all strings in the run have the
character~$c$ in position~$j$.
Intuitively, the runs in $\pr(\ell_q,r_q,j)$ are the subtuples associated
with the children of $q$ if $p(q)$ is set to~$j$.
Let $\last(\ell_q,r_q,j)$ denote the last run in $\pr(\ell_q,r_q,j)$,
and let $\lastj(\ell_q,r_q,j)$ denote the first component of
the triple representing that last run.
For the FA shown in Fig.~\ref{fig:FA},
we have $\pr(1,4,1)=((1,1,\s{a}),(2,2,\s{b}),(3,4,\s{a}))$,
$\last(1,4,1)=(3,4,\s{a})$ and $\lastj(1,4,1)=3$, for example.

Dawson et al.~show that
in an optimal FA it holds that, for any node $v$
associated with a subtuple $\tau(v)=\calS[i,i']$
such that $\com(i,i')\cap \beta(v)\neq \emptyset$,
the part of the FA below $v$ will start with a path with $|\com(i,i')\cap \beta(v)|$
edges whose labels are the letters that all the strings of $\calS[i,i']$
have in the positions in $\com(i,i')\cap \beta(v)$, in arbitrary order~\cite[Property~2]{dawson1996principles}.

\section{The algorithm by Dawson et al.}
\label{sec:dawson}
To aid the understanding of our improved algorithm, we explain in this section
the algorithm by Dawson et al.~\cite{dawson1996principles}
for solving the OFA problem, but using the notation and terminology
of our paper. We refer to their algorithm as the DRSS algorithm.

The DRSS algorithm is based on dynamic programming.
For $1\le i\le i'\le n$,
we define a \emph{FA for the uncommon positions of} $\calS[i,i']$
to be a FA $T'$ for
$\calS[i,i']$ rooted at a node~$v$ but with the assumption that
$T'$ is part of a larger FA~$T$ and the
positions associated with the strict ancestors of~$v$ in $T$
are exactly those in $\com(i,i')$ in arbitrary order,
i.e., $\alpha(v)=\com(i,i')$.
Let $D(i,i')$ denote the minimum size of a FA for the uncommon positions
of $\calS[i,i']$.
For example, for the instance of Fig.~\ref{fig:FA},
we have $D(3,4)=2$ because the subtree rooted at $z$ in that figure
has $2$ edges and
is an optimal FA for the uncommon positions (in this case the
only uncommon position is position~$2$) of $\calS[3,4]=(\s{aab},\s{acb})$.

The key observation by Dawson et al.\ is that the values $D(i,i')$
for $1\le i<i'\le n$ can be computed by dynamic programming via
the following equation:
\begin{align}
    D(i,i') = \min_{k\in \unc(i,i')} \sum_{(j,j',c)\in\pr(i,i',k)} \left(
    |\Delta(i,i',j,j')| + D(j,j')
    \right)
    \label{eq:dawson}
\end{align}
The base case is $D(i,i)=0$ for all $1\le i\le n$. The size of an optimal
FA for $\calS$ can be calculated as $|\com(1,n)|+D(1,n)$. This corresponds to a FA
with the following structure: Starting from its root, there is a path consisting of $|\com(1,n)|+1$ nodes, the first $|\com(1,n)|$ of which are
associated with distinct positions in $\com(1,n)$.
The bottom node of that path is the root of an optimal FA
for the uncommon positions of $\calS=\calS[1,n]$.
Equation~(\ref{eq:dawson}) is correct because it takes the minimum,
over all possible choices of the position $k\in\unc(i,i')$ that could
be used as the label of the root of the FA for the uncommon positions
of $\calS(i,i')$, of the size of the
resulting FA for the uncommon positions of $\calS[i,i']$: The
root of that FA will have $|\pr(i,i',k)|$ children, one
for each triple in $\pr(i,i',k)$. For each triple $(j,j',c)\in\pr(i,i',k)$,
the FA will contain a path starting from the root with $|\Delta(i,i',j,j')|$
edges (to be labeled with the characters
in $\Delta(i,i',j,j')=\com(j,j')\setminus \com(i,i')$),
and the bottom node of that path will be the root
of an optimal FA for the uncommon positions of $\calS[j,j']$.
In the example of Fig.~\ref{fig:FA}, the value of $k$ that
minimizes the expression for $D(1,4)$ is $k=1$, and the
size of the resulting FA is
\begin{align*}
    D(1,4) & = \sum_{(j,j',c)\in\pr(1,4,1)} \left(
    |\Delta(1,4,j,j')| + D(j,j') \right) \\
   & = \left(|\Delta(1,4,1,1)| + D(1,1)\right)
    + \left(|\Delta(1,4,2,2)| + D(2,2)\right)\\
    &  \phantom{=} \mbox{} + \left(|\Delta(1,4,3,4)| + D(3,4)\right)\\
    & = (3+0)+(3+0)+(2+D(3,4))= 10
\end{align*}
Dawson et al.\ analyze the running-time of algorithm DRSS
as follows.
The algorithm computes $O(n^2)$ values $D(i,i')$.
Each such value is calculated as the minimum of at most $m$
expressions, one for each value of $k\in \unc(i,i')$.
For each such expression, $\pr(i,i',k)$ can be determined
in $O(n)$ time. Dawson et al.\ state that the values $|\Delta(i,i',j,j')|$
can be determined for all $(j,j',c)\in \pr(i,i',k)$
together in $O(m)$ time,
after a preprocessing
that requires $O(mn)$ time and space to compute a matrix
that allows one to check in $O(1)$ time for given
$d\in [m]$ and
$1\le j\le j'\le n$ whether all
strings in $\calS(j,j')$ have the same character in
a position~$d$. This gives a bound of
$O(n^2m(n+m))$ on the running-time of the algorithm.
The space usage of their algorithm is $O(n^2+nm)$,
as storing the values $D(i,i')$ requires $O(n^2)$
space and the matrix computed during the preprocessing
takes $O(nm)$ space.

\section{A faster algorithm for optimal factoring automata}
\label{sec:algo}%
We first describe our algorithm for
the OFA problem
in Section~\ref{sec:mainalgo}. Then, in Section~\ref{sec:weighted},
we explain how the algorithm can be adapted to
the weighted OFA problem. In Section~\ref{sec:preproc},
we describe the preprocessing step to construct a data structure
that is used by the main parts of our algorithms
to look up values such as $|\com(i,i')|$
for $1\le i\le i'\le n$ efficiently.

\subsection{Algorithm for the OFA problem}
\label{sec:mainalgo}
The key idea of our faster algorithm for the OFA
problem is to reuse information from the computation
of $D(i,i'-1)$ when calculating $D(i,i')$ in such
a way that $D(i,i')$ can be determined in $O(m)$
time. More precisely, we want to evaluate the
expression
\begin{equation}
    D(i,i',k) = \sum_{(j,j',c)\in\pr(i,i',k)} \left( |\Delta(i,i',j,j')| + D(j,j') \right)
\label{eq:ep1}
\end{equation}
for each $k\in \unc(i,i')$ in $O(1)$ time, so that
$D(i,i')=\min_{k\in \unc(i,i')} D(i,i',k)$ 
(cf.~Equation~(\ref{eq:dawson})) can be obtained
in $O(m)$ time

When going from $D(i,i'-1,k)$ to $D(i,i',k)$, the relevant
changes of~(\ref{eq:ep1}) are: The runs in $\pr(i,i',k)$
differ from the runs in $\pr(i,i'-1,k)$, but not by much:
$\pr(i,i',k)$ can be obtained from $\pr(i,i'-1,k)$
either by adding
the string $S_{i'}$ to the last run in $\pr(i,i'-1,k)$, or
by adding a new run consisting only of $S_{i'}$.
Furthermore, the terms $|\Delta(i,i'-1,j,j')|$ in the sum change to
$|\Delta(i,i',j,j')|$. To handle the latter change without having to
process all terms of the sum separately, we use
the identity $|\Delta(i,i',j,j')|=|\com(j,j')|-|\com(i,i')|$ that
we noted earlier to rewrite Equation~(\ref{eq:ep1})
as follows:
\begin{equation}
\begin{split}
    D(i,i',k) = & \left(\sum_{(j,j',c)\in\pr(i,i',k)} \left( |\com(j,j')| + D(j,j') \right)
	\right)\\
	&- |\pr(i,i',k)| \cdot |\com(i,i')|
\end{split}
\label{eq:ep2}
\end{equation}
In this way, the term $|\com(i,i')|$ that changes when going
from $D(i,i'-1,k)$ to $D(i,i',k)$ is moved outside the sum, and
so a large part of the sum (i.e., all terms except possibly the
one corresponding to the last run) can be reused when determining
$D(i,i',k)$. We now split the expression given for $D(i,i',k)$
in Equation~(\ref{eq:ep2}) into two separate parts as follows:
\begin{align*}
A(i,i',k) &= \sum_{(j,j',c)\in\pr(i,i',k)} \left( |\com(j,j')| + D(j,j') \right)\\
B(i,i',k) &= |\pr(i,i',k)| \cdot |\com(i,i')|
\end{align*}
Note that $D(i,i')=\min_{k\in \unc(i,i')} (A(i,i',k)-B(i,i',k))$.
With suitable bookkeeping and preprocessing, the two factors in
$B(i,i',k)$ can be computed in constant time, as we will show
later. The more challenging task is to compute $A(i,i',k)$ in
constant time, which we tackle next.

Regarding $\pr(i,i',k)$, there are two cases for how it
can be obtained from $\pr(i,i'-1,k)$:
\begin{itemize}
\item Case 1: $S_{i'-1}[k]=S_{i'}[k]$. Let $(j,i'-1,c)=\last(i,i'-1,k)$.
Then $\pr(i,i',k)$ is obtained from $\pr(i,i'-1,k)$ simply be
extending the last run, i.e., by changing the run $(j,i'-1,c)$
to $(j,i',c)$.
We have $\last(i,i',k)=(j,i',c)$ and $\lastj(i,i',k)=\lastj(i,i'-1,k)=j$.
\item Case 2: $S_{i'-1}[k]\neq S_{i'}[k]$. In this case,
$\pr(i,i',k)$ is obtained by taking $\pr(i,i'-1,k)$ and
appending the new run $(i',i',S_{i'}[k])$.
We have $\last(i,i',k)=(i',i',S_{i'}[k])$ and $\lastj(i,i',k)=i'$.
\end{itemize}
In Case~1, we can compute $A(i,i',k)$ from $A(i,i'-1,k)$ as follows:
\begin{equation*}
\begin{split}
A(i,i',k) = & A(i,i'-1,k) - (|\com(j_l,i'-1)|+D(j_l,i'-1)) \\
& + (|\com(j_l,i')|+D(j_l,i'))
\end{split}
\end{equation*}
where $j_l=\lastj(i,i'-1,k)$. This is correct because all terms
of the sum except the final one are the same in $A(i,i'-1,k)$
and $A(i,i',k)$, so it suffices to subtract the final term of the sum
for $A(i,i'-1,k)$ and add the final term of the sum for $A(i,i',k)$.
In Case~2, the formula becomes:
\begin{equation*}
A(i,i',k) = A(i,i'-1,k) + (|\com(i',i')|+D(i',i'))= A(i,i'-1,k)+m
\end{equation*}
\begin{algorithm}[!t]
\caption{Algorithm for the OFA problem}\label{alg:ofa}
\KwData{$n$ strings $S_1,S_2,\ldots,S_n$ of equal length~$m$}
\KwResult{$D(i,i')$ for $1\le i\le i'\le n$}
\For{\label{line:algFor}$i\gets n$ \emph{\textbf{downto}} $1$}{
$D(i,i)\gets 0$\;
\For{$k\gets1$ \emph{\textbf{to}} $m$}{
$p(k)\gets 1$ \Comment*[r]{$p(k)$ is $|\pr(i,i,k)|$}
$l(k)\gets i$ \Comment*[r]{$l(k)$ is $\lastj(i,i,k)$}
$a(k)\gets 0$ \Comment*[r]{$a(k)$ is $A(i,i,k)$}
}
\For{$i'\gets i+1$ \emph{\textbf{to}} $n$}{
  \ForEach{$k\in \unc(i,i')$}{
  \eIf(\tcc*[f]{Case $1$}){$S_{i'-1}[k]=S_{i'}[k]$}{
  \Comment{$p(k)$ and $l(k)$ remain unchanged}
  $a(k)\gets a(k)-(|\com(l(k),i'-1)|+D(l(k),i'-1))
  +(|\com(l(k),i')|+D(l(k),i'))$ \label{line:algD1}\;
  \Comment{$a(k)$ is now $A(i,i',k)$}
  }(\tcc*[f]{Case $2$}){
  $p(k)\gets p(k)+1$ \Comment*[r]{$p(k)$ is $|\pr(i,i',k)|$}
  $l(k)\gets i'$ \Comment*[r]{$l(k)$ is $\lastj(i,i',k)$}
  $a(k)\gets a(k) + m$ \Comment*[r]{$a(k)$ is $A(i,i',k)$}
  }
  }
  $D(i,i')\gets \min_{k\in \unc(i,i')} \left( a(k)-p(k)\cdot |\com(i,i')| \right) $ \label{line:algD} \;
  $k^*(i,i')\gets $ the value of $k$ that yielded the minimum for $D(i,i')$ \label{line:algkstar}\;
  \ForEach{$k \in \com(i,i')$}{
  \Comment{$p(k)$ and $l(k)=i$ remain unchanged}
  $a(k)\gets a(k)\!-\!((|\com(i,i'\!-\!1)|+D(i,i'\!-\!1))
  +(|\com(i,i')|+D(i,i'))$ \label{line:algD2}\;
  }
}
}
\end{algorithm}%
For an efficient implementation, we first create a data
structure that allows us to look up any value $|\com(i,i')|$
for $1\le i\le i'\le n$ in constant time and any set
$\com(i,i')$ or $\unc(i,i')$ in $O(m)$ time. The data
structure also allows us to determine any $\pr(i,i',k)$
in $O(|\pr(i,i',k)|)$ time. The preprocessing
carried out to create this data structure using $O(n^2m)$
time and $O(n^2+nm)$ space is described in
Section~\ref{sec:preproc}.
Furthermore, we maintain arrays $a$, $p$ and $l$
of size $m$ that satisfy the following invariant: At the time
when the algorithm considers the
indices $i$ and~$i'$, the entries in those arrays satisfy
$a(k)=A(i,i',k)$, $p(k)=|\pr(i,i',k)|$
and $l(k)=\lastj(i,i',k)$ for all $1\le k\le m$.
When progressing from a pair $(i,i'-1)$ to a pair
$(i,i')$, the $O(m)$ values in these three arrays can be
updated in constant time per entry, along the lines
discussed above. The resulting algorithm for computing
$D(i,i')$ for all $1\le i\le i'\le n$ in $O(n^2m)$ time
is shown as pseudocode in Algorithm~\ref{alg:ofa}.
In addition to the space $O(nm+n^2)$ used for the
preprocessing, the algorithm uses $O(m)$ space
for the three arrays $p,l,a$ and $O(n^2)$ space for~$D$,
so the total space usage is $O(mn+n^2)$.
Note that for each pair $(i,i')$ with $i'>i$
the algorithm processes the values of $k$
in $\unc(i,i')$ before those in $\com(i,i')$.
This is important because the formula
for updating $a(k)$ for $k\in \com(i,i')$
uses the value $D(i,i')$ (see Line~\ref{line:algD2}), but that value
can only be calculated (see Line~\ref{line:algD}) once $a(k)$ has
been determined for all $k\in \unc(i,i')$.
Note that $l(k)=\lastj(i,i',k)>i$ in Line~\ref{line:algD1}
as $\pr(i,i',k)$ contains at least two runs if
$k\in \unc(i,i')$, so the values $D(l(k),i'-1)$
and $D(l(k),i')$ accessed in  Line~\ref{line:algD1}
have both been computed already (as the for-loop in
Line~\ref{line:algFor} iterates through the values of $i$
in decreasing order).

When the values $D(i,i')$ (together with the values $k^*(i,i')$ that
represent the choice of $k$ that yields the minimum in the formula
for $D(i,i')$, see Line~\ref{line:algkstar}) have been calculated using
Algorithm~\ref{alg:ofa}, the size of the optimal
FA is given by $|\com(1,n)|+D(1,n)$.
To construct the optimal FA itself, we proceed
as follows: Create a root node~$v_1$. Let $x=|\com(1,n)|$.
If $x>0$,
create a path $(v_1,v_2,v_3,\ldots,v_{x+1})$ such
that the nodes $v_1,\ldots,v_x$ are labeled with the
positions in $\com(1,n)$ and the downward edges
of these nodes are labeled with the characters
in those positions of the strings in~$\calS$.
Label $v_{x+1}$ with $p(v_{x+1})=k^*(1,n)$,
All the nodes $v_j$ with $1\le j\le x+1$
have $\tau(v_j)=\calS$.
Create a downward edge from $v_{x+1}$ with label $c$
for each run $(j,j',c)$ in $\pr(1,n,k^*(1,n))$.
Note that the child node $w$ at the bottom of a downward
edge for run $(j,j',c)$ has $\tau(w)=\calS[j,j']$.
For any child node $w$ with $\tau(w)=\calS[j,j']$ for $j<j'$
of a parent node $v$ with $\tau(v)=\calS[i,i']$,
construct the FA rooted at $w$ analogously:
Start with a path of $\Delta(i,i',j,j')$
edges (the first of which is the edge from $v$ to $w$ and is given label $S_j[k^*(i,i')]$),
assign label $p(z)=k^*(j,j')$ to
the node $z$ at the end of that path,
and create a downward edge from $z$
with label $c$ for each run $(r,r',c)$ in $\pr(j,j',k^*(j,j'))$.
The recursion stops at depth $m$, i.e., when the
nodes created are leaves of the optimal FA.

The data structure described in Section~\ref{sec:preproc}
also allows us to iterate over the runs in $pr(i,i',k)$,
for any given values $i$, $i'$, and~$k$, in time $O(|\pr(i,i',k)|)$.
For a node $v$ in the optimal FA that is a leaf or has more than
one child, let $w$ be the node of
maximum depth among all strict ancestors of $v$
that have more than one child (or the root if no
strict ancestor of $v$ has more than one child).
Assume that $\tau(v)=\calS[j,j']$ and $\tau[w]=\calS[i,i']$.
In our construction of the optimal FA,
it takes $O(m)$ time to determine $\Delta(i,i',j,j')$
and create the path from $w$ to~$v$.
If $v$ is not a leaf, it takes $O(|\pr(j,j',k^*(j,j'))|)$
time to create the $|\pr(j,j',k^*(j,j'))|$ children of~$v$.
Adding up these times over all the $O(n)$ nodes
$v$ that have more than one child or are leaves,
the total time for constructing paths is $O(mn)$
and the total time for creating children of nodes
with more than one child is $O(n)$. Thus, once
the values $D(i,i')$ and $k^*(i,i')$ have been
computed using Algorithm~\ref{alg:ofa},
the optimal FA can be constructed in $O(nm)$ time.
\FloatBarrier

Thus, we obtain the following theorem.

\begin{theorem}
The OFA problem can be solved in $O(n^2m)$ time
using $O(nm+n^2)$ space.
\end{theorem}

\subsection{Adaptation to the weighted OFA problem}
\label{sec:weighted}
Recall that, in the weighted OFA problem,
the cost of a non-leaf node with label $k$
that has at least two children
is $\cchoice(k)$ and the cost of an edge with label $c$ whose parent node $v$ has
label $p(v)=k$ is $\cunify(k,c)$.
With $D_w(i,i')$ denoting the minimum cost of a FA
for the uncommon positions of $\calS[i,i']$,
Equation~(\ref{eq:dawson}) can be adapted to the weighted
case as follows \cite{dawson1996principles}:
\begin{align*}
    D_w(i,i') = \min_{k\in \unc(i,i')} \cchoice(k) + \sum_{(j,j',c)\in\pr(i,i',k)} \left(
    \Delta_w(i,i',j,j') + D_w(j,j')
    \right)\,,
\end{align*}
where $\Delta_w(i,i',j,j') = \sum_{h \in \Delta(i,i',j,j')} \cunify(h,S_j[h])$.
We again consider the terms for each value of $k$ separately:
\begin{equation}
    D_w(i,i',k) = \cchoice(k) + \sum_{(j,j',c)\in\pr(i,i',k)} \left( \Delta_w(i,i',j,j') + D_w(j,j') \right)
	\label{eq:epweighted}
\end{equation}
If we let $\com_w(i,i')=\sum_{h\in \com(i,i')} \cunify(h,S_j[h])$,
we have
$\Delta_w(i,i',j,j') = \com_w(j,j')-\com_w(i,i')$
and can rewrite (\ref{eq:epweighted}) as follows (analogously
to~(\ref{eq:ep2})):
\begin{equation*}
\begin{split}
    D_w(i,i',k) = & \cchoice(k) + \left(\sum_{(j,j',c)\in\pr(i,i',k)} \left( \com_w(j,j') + D_w(j,j') \right)
	\right)\\
	&- |\pr(i,i',k)| \cdot \com_w(i,i')
\end{split}
\end{equation*}
We define 
$$
A_w(i,i',k) = \sum_{(j,j',c)\in\pr(i,i',k)} \left( \com_w(j,j') + D_w(j,j') \right)
$$
and can compute these values analogously: In Algorithm~\ref{alg:ofa},
we change Line~\ref{line:algD1}
to
$$
a(k) \gets a(k) - (\com_w(l(k),i'-1) + D_w(l(k),i'-1)) + (\com_w(l(k),i')+D_w(l(k),i'))
$$
and Line~\ref{line:algD2} to
$$
a(k) \gets a(k) + \com_w(i',i')
$$
and the computation of $D(i,i')$ in Line~\ref{line:algD}
becomes:
$$
D_w(i,i') = \min_{k\in \unc(i,i')} \left(\cchoice(k) + a(k) - |\pr(i,i',k)|\cdot \com_w(i,i')\right)
$$
As also shown in the following section,
the data structure constructed in the preprocessing can be
extended (without affecting the asymptotic time and space bounds)
so that it can be used to determine $\com_w(i,i')$ in constant time
for any $1\le i\le i'\le n$ as well.

\subsection{Preprocessing}
\label{sec:preproc}
For a given string tuple $\calS=(S_1,\ldots,S_n)$ with
$n$ strings of length~$m$ each, we want to compute
a data structure that allows us to determine
$|\com(i,i')|$ or $\com_w(i,i')$ in constant time, to compute
$\com(i,i')$ or $\unc(i,i')$ in $O(m)$ time,
and to compute $\pr(i,i',k)$ in
$O(|\pr(i,i',k)|)$ time.

First, we create an $n\times m$ matrix $R$
such that $R(i,k)$ is the number of consecutive
strings in $\calS$, starting with $S_i$, that
have the same character in position~$k$.
Formally, $R(i,k)=\max\{\ell \mid S_i[k]=S_{i+1}[k]=\cdots=S_{i+\ell-1}[k]\}$.
$R$~can be computed in $O(nm)$ time by
setting $R(n,k)=1$ for all $k$ and
using the equation
$$
R(i,k)=\left\{ \begin{array}{ll}
R(i+1,k)+1 & \mbox{if $S_i[k]=S_{i+1}[k]$}\\
1 & \mbox{if $S_i[k]\neq S_{i+1}[k]$}
\end{array}
\right.
$$
for $1\le i\le n-1$ (in order of decreasing~$i$) and $1\le k\le m$.

Now, compute an $n\times n$ matrix $C$
by setting $C(i,i') = |\{k \in [m] \mid R(i,k)\ge i'-i+1 \}|$
for $1\le i\le i'\le n$. $C(i,i')$ contains
the number of positions $k$ such that there
are at least $i'-i+1$ consecutive strings,
starting with $S_i$, in $\calS$ that have
the same character in position~$k$. This
shows that $C(i,i')=|\com(i,i')|$.
The computation of $C$ takes time $O(n^2m)$.

If we want to handle the weighted OFA
problem, we additionally (or instead of~$C$)
compute an $n\times n$ matrix $C_w$
by setting
$$C_w(i,i')=\sum_{\substack{k \in [m] \\
R(i,k)\ge i'-i+1}} \cunify(k,S_i[k])\;.$$
Each entry of $C_w$ can be computed in
$O(m)$ time, and we have $C_w(i,i')=\com_w(i,i')$.

The computation of $R$, $C$ and/or $C_w$
takes $O(n^m)$ time and $O(nm+n^2)$ space.

Once $R$ and $C$ and/or $C_w$ have been computed,
queries can be answered as follows:
To determine $|\com(i,i')|$ in constant time, we
return $C(i,i')$.
To determine $|\com_w(i,i')|$ in constant time, we
return $C_w(i,i')$.
To list the positions
in $\com(i,i')$ in $O(m)$ time, we
check for each position $k\in [m]$ whether it satisfies
$R(i,k)\ge i'-i+1$, and return the positions that meet
this condition. For $\unc(i,i')$, we change the condition
to $R(i,k)<i'-i+1$.
To list the runs in $\pr(i,i',k)$ in $O(|\pr(i,i',k)|)$
time, we proceed as follows:
The first run is $(i,\min\{i+R(i,k)-1,i'\},S_i[k])$.
Once a run $(j,j',c)$ with $j'<i'$ has been determined, the
run following it is $(j'+1,\min\{j'+R(j'+1,k),i'\},S_{j'+1}[k])$.
Therefore, each run in $\pr(i,i',k)$
can be determined in constant time.

\section{Conclusions}
\label{sec:conc}%
In this paper, we have given an algorithm that solves the
OFA problem for~$n$ given strings of equal length~$m$
in $O(n^2m)$ time and $O(n(n+m))$ space.
The algorithm can be adapted to the weighted OFA problem
in the same time and space bounds. The running-time of our
algorithm is better than that of the previously known
algorithm by Dawson et al.~\cite{dawson1996principles}
by a factor of $n+m$. The main idea leading to the improvement
is reusing information from previously computed entries of
the dynamic programming table when computing new entries.

Our algorithms may be parallelizable using methods such as
exploiting table-parallelism~\cite{freire1995exploiting}, which has been
used to parallelize the previously known algorithm for the
OFA problem.

\bibliographystyle{plain}
\bibliography{b}
\end{document}